\documentclass[aps,prl,reprint,floatfix,preprintnumbers,showpacs,citeautoscript,
superscriptaddress]{revtex4-1}
\usepackage{graphicx}
\usepackage{amsfonts,amssymb,amsmath}
\usepackage{mathtools}
\usepackage{suffix}
\usepackage[utf8x]{inputenc}
\usepackage{xcolor}
\usepackage{etoolbox}

\apptocmd{\sloppy}{\hbadness 10000\relax}{}{}
\let\originalleft\left
\let\originalright\right
\def\left#1{\mathopen{}\originalleft#1}
\def\right#1{\originalright#1\mathclose{}}
\newcommand{\bo}{\mathbf}
\newcommand{\bs}{\boldsymbol}

\newcommand{\ra}{\rightarrow}

\DeclareMathOperator{\sgn}{sgn}
\DeclareMathOperator{\Tr}{\textrm{Tr}}

\DeclarePairedDelimiterX\MeijerM[3]{\lparen}{\rparen}%
{\begin{smallmatrix}#1 \\ #2\end{smallmatrix}\delimsize\vert\,#3}

\newcommand\MeijerG[8][]{%
  G^{\,#2,#3}_{#4,#5}\MeijerM[#1]{#6}{#7}{#8}}

\WithSuffix\newcommand\MeijerG*[7]{%
  G^{\,#1,#2}_{#3,#4}\MeijerM*{#5}{#6}{#7}}

\begin{document}

\title{Hybridization and anisotropy in the exchange interaction in 3D Dirac semimetals}
\date{\today}

\author{D.\ Mastrogiuseppe}
\affiliation{Department of Physics and Astronomy, and Nanoscale and Quantum
Phenomena Institute, \\ Ohio University, Athens, Ohio 45701--2979, USA}
\affiliation{Instituto de F\'isica Rosario (CONICET), 2000 Rosario, Argentina}
\author{N.\ Sandler}
\author{S.\ E.\ Ulloa}
\affiliation{Department of Physics and Astronomy, and Nanoscale and Quantum
Phenomena Institute, \\ Ohio University, Athens, Ohio 45701--2979, USA}

\begin{abstract}
We study the Ruderman-Kittel-Kasuya-Yosida interaction in 3D Dirac semimetals.  Using retarded Green's functions in real space, we obtain and analyze asymptotic expressions for the interaction, with magnetic impurities at different distances and relative angle with respect to high symmetry directions on the lattice. We show that the Fermi velocity anisotropy in these materials produces a strong renormalization of the magnitude of the interaction, as well as a correction to the frequency of oscillation in real space. Hybridization of the impurities to different conduction electron orbitals are shown to result in interesting anisotropic spin-spin interactions which can generate spiral spin structures in doped samples.
\end{abstract}

\pacs{71.55.Ak, 75.30.Hx, 75.10.Lp, 75.25.Dk}

\maketitle

\emph{Introduction}.  Dirac semimetals are fascinating new materials that can be considered analogues of graphene in three dimensions. They possess robust Dirac points that are protected by crystalline symmetry, and strong spin-orbit interaction (SOI). Na$_3$Bi and Cd$_3$As$_2$ are among these compounds, where the unconventional Dirac character was detected in angle resolved photoemission and transport experiments \cite{Liu2014Na, Kushwaha2015, Liu2014Cd, Yi2014, Neupane2014}. Many more materials have been proposed as promising candidates  \cite{Gibson2015}.
When time-reversal and/or inversion symmetry is broken, the degeneracy of each Dirac cone splits without the opening of a gap, leading to the Weyl semimetal phase. The latter phase is characterized by unconventional properties, such as a chiral anomaly and Fermi arcs on the surfaces, as recently measured in TaAs  \cite{Xu2015Ta, Lv2015Ta1, Lv2015Ta2}, NbAs \cite{Xu2015NbAs}, and NbP \cite{Fei2015NbP}. 
These unusual properties suggest that magnetic impurities can reveal exotic behavior, as predicted, for instance, for the Kondo effect \cite{Principi2015, Mitchell2015, Sun2015}.

Impurities are ubiquitous in the preparation of experimental samples and they can also be purposely introduced by different processes.
It is well known that in metallic hosts, magnetic impurities interact effectively through the electron gas, and that this interaction has an oscillatory decay  when the separation  between them is increased. This  Ruderman-Kittel-Kasuya-Yosida (RKKY) interaction \cite{Ruderman1954, Kasuya1956, Yosida1957} gets more complicated when the host material has a more involved band structure and/or additional degrees of freedom. For instance, graphene is predicted to have an unconventional decay dependence for the charge neutral case \cite{Saremi2007, Sherafati2011}. Strong SOI can also affect the behavior, giving rise to spin-spin interactions that contain anisotropic terms such as Ising and Dzyaloshinskii-Moriya (DM) interactions on top of the usual Heisenberg-like terms \cite{Imamura2004}. 

In this work we study the RKKY interaction in 3D Dirac semimetals, focusing on Na$_3$Bi and Cd$_3$As$_2$, two compounds with strong Fermi velocity anisotropy \cite{Liu2014Na, Liu2014Cd}. Starting with a low energy model, we consider magnetic impurities that hybridize with Na-$s$ and Bi-$p$ orbitals, the most relevant near the band crossings that build the Dirac points \cite{Wang2012}. We obtain asymptotic expressions for the interaction, and analyze its behavior as a function of the impurity separation as related to the underlying lattice. The role of the SOI in these materials manifests uniquely when each impurity hybridizes with a different conduction electron orbital, resulting in strong interaction anisotropies. We also show that Dirac dispersion anisotropies seen in these materials have strong impact on the amplitude and spatial dependence of the effective exchange interaction.

\emph{Model}.
Two magnetic impurities coupled to an electron gas can be described by the 
Hamiltonian
\begin{equation} \label{eq:imp_Hamilt}
 H = H_0 + J \sum_{j=1,2} \bs{S}_j \cdot \bs{s}(\bs R_j),
\end{equation}
where $H_0$ is the unperturbed Hamiltonian for the host material, $\bs s(\bs 
r) =  \sum_{i} \delta(\bs r - \bs r_i)\bs\sigma_i$, in units of $\frac{\hbar}{2}$,  is the spin density operator for the conduction electrons, where $\bs r_i$ and $\bs\sigma_i$ are the position and Pauli matrices for electron $i$. $\bs S_j$ is the localized spin operator for impurity $j$. At second order in perturbation theory in the interaction parameter $J$, one can obtain an effective Hamiltonian that describes the carrier mediated interaction between the impurities separated by a distance vector $\bs R$ 
\begin{equation} \label{eq:rkky_hamilt}
 H_{\text{RKKY}} = J^2 \sum_{\mu,\mu'} S_1^{\mu}\, 
\chi_{\mu,\mu'}(\bo R)\, S_2^{\mu'},
\end{equation}
where $\chi_{\mu,\mu'}$ is the static spin susceptibility tensor of the
electron gas, and $\mu,\mu'$ represent the Cartesian components \cite{Mattis}.
For conventional electron gases, and in the absence of SOI, the susceptibility tensor is diagonal so that the effective spin-spin coupling is isotropic. Moreover, the interaction decays as $|\bo R|^{-D}$, where $D$ is the dimensionality of the system \cite{Aristov1997}. When the SOI is present, anisotropic components of Ising and/or DM type may appear \cite{Imamura2004}. Additionally, the presence of particular features in the band structure, such as Dirac points, may change the decay exponent (e.g. in graphene, $|\textbf{R}|^{-3}$ at the Dirac point \cite{Saremi2007, Sherafati2011}). 

A convenient way to calculate the $T=0$ spin susceptibility for a system with SOI is via the real space retarded Green's functions \cite{Imamura2004},
\begin{equation}\label{eq:suscept}
\begin{split}
\chi_{\mu,\mu'}^{\alpha,\beta}(\bo R) = & -\frac{1}{\pi} \textrm{Im}\Tr 
\int_{-\infty}^{\omega_F} \sigma_\mu  G^{\alpha,\beta}(\bo R,\omega^+)\\
&\times
\sigma_{\mu'} G^{\beta,\alpha}(-\bo R,\omega^+) d\omega,
\end{split}
\end{equation}
where $\omega^+=\omega+i 0^+$, $\omega_F$ is the Fermi energy, and the trace is over spin components.
$\alpha$ and $\beta$ denote sets of additional degrees of freedom (other than 
spin) that characterize the host.

\begin{figure}[htb]
	\includegraphics[width=0.45\textwidth]{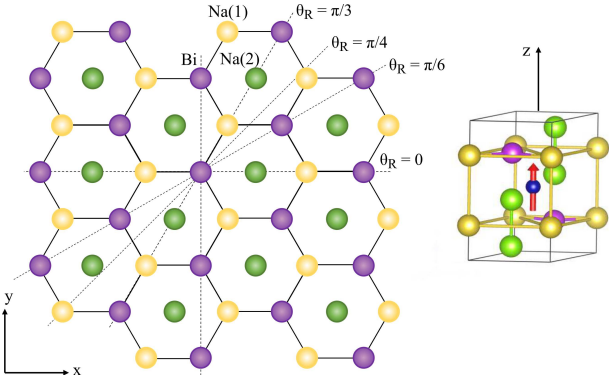}
	\caption{\label{fig:lattice} (Color online) Lattice structure of Na$_3$Bi in the $xy$ plane (left) with dashed lines denoting high symmetry directions. Na$_3$Bi unit cell \cite{Vesta} with a magnetic impurity in its center as a possible location (right).}
\end{figure}

Figure \ref{fig:lattice} shows the hexagonal Na$_3$Bi lattice structure in the $xy$ plane and its unit cell. A low energy Hamiltonian was proposed in Ref.\ \onlinecite{Wang2012} for Na$_3$Bi.
This Hamiltonian also describes Cd$_3$As$_2$ with an appropriate set of parameters. Up to second order in momentum, it reads 
\begin{align}
 H = \epsilon_0(\bs k)\tau_0 \sigma_0 + M(\bs k) \tau_z \sigma_0 + A (k_x 
\tau_x \sigma_z - k_y \tau_y \sigma_0),
\end{align}
where $\epsilon_0(\bs k) = C_0 + C_1 k_z^2 + C_2 (k_x^2+ k_y^2)$, 
$M(\bs k) = M_0 - M_1 k_z^2 - M_2 (k_x^2+ k_y^2)$, and  $C_i$, $M_i$, $A$ are 
parameters that depend on the specific material \footnote{For Na$_3$Bi, the values of the parameters are \cite{Wang2012}: $C_0 \simeq$ -0.06 eV, $C_1 \simeq$ 8.75 eV\AA$^2$, $C_2\simeq$ -8.4  eV\AA$^2$, $M_0\simeq$ -0.09 eV, $M_1\simeq$ -10.64 eV\AA$^2$, $M_2\simeq$ -10.36  eV\AA$^2$, $A\simeq$ 2.46  eV\AA.}.
In the case of Na$_3$Bi, the Hamiltonian is expressed in the basis of relevant orbitals around the linear band crossings: $\left(|S,\frac12\rangle, |P,\frac32\rangle, |S,-\frac12\rangle, |P,-\frac32\rangle\right)$, where $S$ stands for Na-$3s$ bonding orbitals and $P$ for Bi-$6p$ antibonding orbitals \cite{Wang2012}. The second quantum number in the kets indicates the $z$ projection of the total angular momentum, upon consideration of the SOI. Notice that the most relevant $p$-like states near the Dirac points correspond to $j=\frac32$ and $m_j = \pm\frac32$, 
where $j(j+1)$ and $m_j$ are eigenvalues of total angular momentum operators 
$\hat J^2$ and $\hat J_z$ respectively. The Pauli matrices $\tau$ and $\sigma$ act in the orbital and spin spaces respectively.
There are two Dirac points at $\bs K^{\pm 1}= (0,0,\pm \sqrt{M_0/M_1})$, 
protected by the crystalline symmetry.
One can expand the Hamiltonian around these two points to get an effective 
low-energy model. In dimensionless form
\begin{align}
 H(\bs \kappa) = \lambda q_z \nu_z \tau_z \sigma_0 + \nu_0(k_x \tau_x \sigma_z - k_y \tau_y \sigma_0),
\end{align}
where $\bs \kappa \equiv (k_x, k_y, q_z)$, the energy is expressed in terms of $A/a= 0.451\, \text{eV}$ for Na$_3$Bi; in what follows, all the energies will be expressed in this scale. $k_x, k_y$ are in units of the inverse lattice spacing $1/a$ ($a\simeq 5.45 \text{\AA}$), $q_z$ is the momentum in the $z$ direction, measured from the Dirac points and in units of $1/c$ ($c \simeq 9.65 \text{\AA}$). The factor $\lambda\simeq 0.25$ characterizes the Fermi velocity anisotropy in the $z$ direction \cite{Liu2014Na} ($\lambda \simeq 0.25$ for Cd$_3$As$_2$ as well \cite{Liu2014Cd}). The Pauli matrices $\nu$ operate in the valley degree of freedom.

From this Hamiltonian we obtain the Green's function matrix in momentum 
space $G(\bs \kappa, \omega) = [\omega^+ -H(\bs \kappa)]^{-1}$. In the present case, $G$ is an $8\times 8$ matrix containing orbital, spin, and valley degrees of freedom. This matrix is block diagonal, and the inversion is simply calculated as an inversion of several $2\times 2$ blocks. One gets
\begin{equation}
 G(\bs \kappa, \omega) = \rho(\bs \kappa, \omega)^{-1} \left[\omega^+ + H(\bs \kappa)\right],
\end{equation}
where $\rho(\bs \kappa, \omega) \equiv \omega_+^2 - k_x^2-k_y^2 - \lambda^2 q_z^2$.
It is convenient to separate the Green's function in terms of the only two spin matrices in $H$, as \cite{Imamura2004}
\begin{equation}
 G(\bs\kappa) = G_0(\bs\kappa) \sigma_0 + G_z(\bs\kappa) \sigma_z,
\end{equation}
with
\begin{align}
 G_0(\bs\kappa) &= \rho(\bs \kappa, \omega)^{-1} \left[\omega_+ 
+\lambda q_z \nu_z \tau_z - k \sin\theta_k \nu_0 \tau_y\right],\\
G_z(\bs\kappa) &= \rho(\bs \kappa, \omega)^{-1} k \cos\theta_k \nu_0\tau_x,
\end{align}
where we have introduced cylindrical coordinates, $k = (k_x^2 + k_y^2)^{\frac12}$, $\theta_k = \arctan\left(k_y/k_x\right)$.
One can make further advances in determining the terms  generated by the trace operation. As the Fourier transform does not change the spin character of the Green's function,  Eq.\ \eqref{eq:suscept} will have terms 
of the form $\Tr \left[G_0^{\alpha,\beta}(\bs R)\sigma_\mu + 
G_z^{\alpha,\beta}(\bs R)\sigma_\mu\sigma_z\right] \times\left[G_0^{\beta,\alpha}(-\bs R)\sigma_{\mu'} + G_z^{\beta,\alpha}(-\bs R)\sigma_{\mu'}\sigma_z\right]$. Then, we can write $\chi_{\mu,\mu'}^{\alpha,\beta} = -\frac{2}{\pi} 
\textrm{Im}\int_{-\infty}^{\omega_F} A_{\mu,\mu'}^{\alpha,\beta}\: d\omega$, where
\begin{align}
 A_{x,x}^{\alpha,\beta} & = G_0^{\alpha,\beta}(\bs R)G_0^{\beta,\alpha}(-\bs R)- G_z^{\alpha,\beta}(\bs R)G_z^{\beta,\alpha}(-\bs 
R),\nonumber\\
A_{z,z}^{\alpha,\beta} &= G_0^{\alpha,\beta}(\bs R)G_0^{\beta,\alpha}(-\bs R)+G_z^{\alpha,\beta}(\bs R)G_z^{\beta,\alpha}(-\bs R),\\
A_{x,y}^{\alpha,\beta} &=  i G_0^{\alpha,\beta}(\bs R)G_z^{\beta,\alpha}(-\bs R)-i G_z^{\alpha,\beta}(\bs R)G_0^{\beta,\alpha}(-\bs R),\nonumber
\end{align}
with $A_{y,y} = A_{x,x}$, $A_{y,x} =-A_{x,y}$, and the remaining cross terms vanish.
Using these expressions in Eq.\ \eqref{eq:rkky_hamilt}, one gets in-plane XX ($x, x$),  Ising ($z, z$), and DM ($x, y$) components
\begin{equation}\label{eq:rkky_components}
 \begin{split}
  H_{RKKY} =& J^2 \big[\chi_{x,x} (S_1^x S_2^x + S_1^y S_2^y) + \chi_{z,z} S_1^z S_2^z \\
  &+ \chi_{x,y}(\bs S_1\times\bs S_2)_z\big],
 \end{split}
\end{equation}
as expected when SOI is present \cite{Imamura2004, Lyu2007, Schulz2009, Biswas2010, Klinovaja2013, Parhizgar2013, Mastrogiuseppe2014}.
The appearance of each component depends on the coupling of each 
impurity to the different orbital and valley degrees of freedom. There is no reason to couple inequivalently to each valley, and the susceptibility contains Green's functions which are the sum of each valley components: $G(\bs R) = \sum_\nu G^\nu(\bs R)$. 
The products of Green's functions will generate intra- and inter-valley terms in the susceptibility due to the scattering of the conduction electrons with the localized impurities. 

From the Hamiltonian matrix in momentum space, one can see that $S$ and $P$ orbitals are connected by the propagators because the SOI mixes them. In particular, we see that $G_z(\bs\kappa)$ contains only $\tau_x$, so the propagator does not connect $S$ and $P$ orbitals to themselves: $G_z^{\nu,S,S}=G_z^{\nu,P,P}=0$. This implies that the in-plane and Ising terms  in Eq.\ \eqref{eq:rkky_components} are equal and that the DM term vanishes.
Therefore if both impurities are coupled only to either $S$ or $P$ orbitals, the RKKY interaction will be of completely isotropic (Heisenberg), as expected for a host without SOI.

The appearance of anisotropic interactions between impurities requires one of them to be connected to an $S$ orbital and the other to a $P$ orbital. This does not require each impurity to be coupled to only one type of orbital. In fact, a probable impurity position would be in the middle of the tetragonal unit cell of the material (see Fig.\ \ref{fig:lattice}), which would (locally) preserve the inversion symmetry. For impurities located at this high-symmetry point, it is expected that they would connect to both $S$ and $P$ orbitals, so that the effective interaction in \eqref{eq:rkky_components} will have all three terms.
Although the analysis of possible locations and orbital configurations of the local magnetic moments is beyond the scope of this paper, in the following we analyze the possible diagonal and non-diagonal orbital components for the different interactions.

\begin{figure}[htb]
	\includegraphics[width=0.45\textwidth]{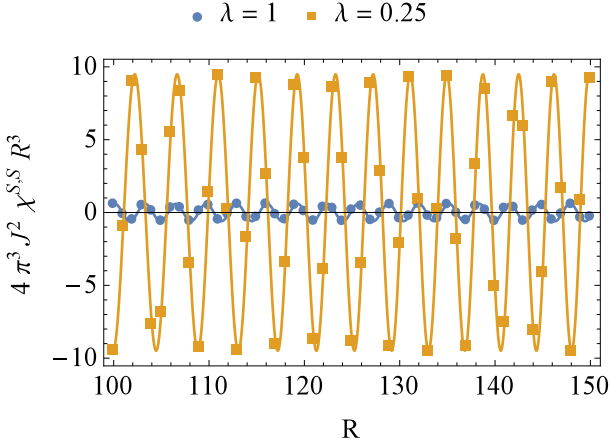}
	\caption{\label{fig:Jeff_vs_R} (Color online) Effective impurity interaction as a function of their separation $R$ in the $xy$ plane. The anisotropy in the Fermi velocity, characterized by $\lambda=0.25$, has a big impact on the strength of the interaction with respect to the isotropic case ($\lambda=1$). It also introduces a correction in the period of the oscillation.}
\end{figure}

\emph{Diagonal orbital components.}
When both impurities connect to the same type of orbital, we have that $G_0^{\nu=1, S, S}(\bs\kappa) = G_0^{\nu=-1, P, P}(\bs\kappa) = \rho(\bs \kappa, \omega)^{-1} (\omega_+ + \lambda q_z)$, and $G_0^{\nu=-1, S, S}(\bs\kappa) = G_0^{\nu=1, P, P}(\bs\kappa) = \rho(\bs \kappa, \omega)^{-1} (\omega_+ - \lambda q_z)$. The real space version, after integration on $\theta_k$ and $k$ [in the $(0,\infty)$ range, valid for large impurity separations], can be written as
\begin{equation} \label{eq:g0ss}
 \begin{split}
  G_0^{\nu,S,S}&(\bo R) = -\frac{e^{i \nu K_z R_z}}{(2\pi)^2}\!\int_{-\infty}^\infty  e^{i q_z 
	R_z}\\
&\times (\omega_+ +\nu \lambda q_z) K_0\left(\!\!R \sqrt{\lambda^2 
	q_z^2 - \omega_+^2}\right) dq_z,
\end{split}
\end{equation}
where $\bo R$ is in cylindrical coordinates, $R$ is the radial coordinate in the $xy$ plane, and $K_0$ is the Bessel function. $k_z$ has been replaced by $\pm K_z + q_z$. The analytic continuation $\omega^+ \rightarrow \omega$ and  the branch cut in the square root allow one to write 
\begin{equation}
 G_0^{\nu,S,S}(\bo R) = -\frac{e^{i \nu K_z R_z}}{(2\pi)^2}\left(I_0- i\sgn(\omega) I_1\right),
\end{equation}
where
\begin{equation}
 \begin{split}
  I_0 &= \left(\int_{-\infty}^{-\frac{|\omega|}{\lambda}}+ 
\int_{\frac{|\omega|}{\lambda}}^{\infty} \right)  e^{i q_z 
R_z} (\omega +\nu \lambda q_z) K_0\left(u\right)\;dq_z,\\
  I_1 &=  \int_{-\frac{|\omega|}{\lambda}}^{\frac{|\omega|}{\lambda}}
e^{i q_z R_z} (\omega +\nu \lambda q_z)
 K_0\left(\!-i\sgn(\omega) v\right) \;dq_z,
\end{split}
\end{equation}
with $u = R \sqrt{\lambda^2 q_z^2 - \omega^2}$ and $v =  R \sqrt{\omega^2 -\lambda^2 q_z^2}$. 
Lacking analytical solutions, we proceed with the case $R \gg R_z$, which allows one to obtain asymptotic expressions.
Considering the case where the Fermi energy lies below the Dirac points,  $\omega < 0$, and adding the contributions of the two valleys, one gets \cite{suppl}
\begin{equation}
\begin{split}
G_0^{S,S}&(\bo R, \omega) \simeq
 -\frac{1}{\pi^2 \lambda R^2} \bigg(e^{\frac{i 3\pi}{4}}\cos\left(\!K_z R_z - \frac{|\omega| R_z}{\lambda}\right)\\
& + i \frac{\pi \omega R}{2} \exp\left(\! i R \omega\left[1+\frac{R_z^2}{2\lambda^2 R^2}\right]\right)\\
&\times\left[\cos(K_z R_z) + i\frac{R_z}{\lambda R} \sin(K_z R_z)\right]\bigg),
\end{split}
\end{equation}
with the same expression for $G_0^{P,P}(\bo R, \omega)$.
We can now calculate the susceptibility, by integrating over $\omega$. The integration generates many terms, with the most relevant in the $R$ asymptotic limit  given by \cite{suppl}
\begin{equation} \label{chi_diag}
\begin{split}
\chi^{S,S}_{x,x}&(\bo R,\omega_F) = \chi^{P, P}_{z,z}(\bo R,\omega_F)\\
&\simeq -\frac{\omega_F^2}{4 \pi^3\lambda^2 R^3}\cos^2(K_z R_z)
\cos\left(2 R\left[1 + \frac{R_z^2}{2 \lambda^2 R^2}\right] \omega_F\right).
\end{split}
\end{equation}
Notice there is no angular dependence.
These effective in-plane spin-spin interactions decay as $1/R^3$, while there is no decay for separations in the $z$ direction; they only oscillate with $R_z$ ($\ll R$). There are, however, important corrections due to the dispersion anisotropy. The form of the spatial term inside the second cosine  comes from a second order expansion of an effective distance given by $\hat{R} \equiv \sqrt{R^2 + \lambda^2 R_z^2} \simeq R\left[1 + \frac{R_z^2}{2 \lambda^2 R^2}\right]$. For $\lambda = 1$, which corresponds to a completely isotropic Fermi velocity, we recover the expected isotropic distance dependence in 3D. 
Another important effect of the anisotropy is to modulate the amplitude of the interaction. It decreases for $\lambda>1$, with respect to the isotropic case. For  materials with $\lambda<1$, such as Na$_3$Bi and Cd$_3$As$_2$, the interaction is significantly enhanced (Fig.\ \ref{fig:Jeff_vs_R}).
Notice that the  interaction decays quadratically in energy towards the Dirac point.
The interesting oscillatory (and always positive) term that comes from inter-valley scattering, modulates the usual oscillatory RKKY term. When $K_z R_z$ is an odd multiple of $\frac{\pi}{2}$, the interaction vanishes for any value of $R$ or band filling. In Na$_3$Bi, where $K_z\simeq 0.82\times \frac1c$, this will happen for $R_z \simeq 3.83 \left(n-\frac12\right)c$, where $n$ is an integer.
Exactly at the Dirac nodes, $\omega_F=0$, $\chi^{S,S}$ in Eq.\ \eqref{chi_diag} vanishes at the third order in the asymptotic expansion in $R$. At the next order in the expansion, one gets no oscillation with the in-plane distance $R$, and $\sim R^{-4}$ decay.

\emph{Off-diagonal orbital components.}
Now we consider the case in which one impurity is connected to an $S$ orbital and the second one to a $P$ orbital. The Green's functions have the following properties: $G_0^{\nu,S,P}(\bs\kappa,\omega)= -G_0^{\nu,P,S}(\bs\kappa,\omega) = i\rho(\bs \kappa, \omega)^{-1} k \sin\theta_k$. 
Proceeding in a similar way as in the diagonal case, one gets \cite{suppl}
\begin{align}
G_0^{S,P}(\bo R,\omega) &\simeq  \frac{\sin(\theta_R)}{2\pi^2} \cos\left( 
K_z R_z\right)\: f(R, R_z, \omega),\\
G_z^{S,P}(\bo R,\omega) &\simeq  \frac{i \cos(\theta_R)}{2\pi^2}\cos\left( 
K_z R_z\right)\: f(R, R_z, \omega),
\end{align}
where
\begin{equation}
\begin{split}
f(R, &R_z, \omega) = -\frac{1}{\lambda R^2} \left[\frac{4}{R \omega}\left(1+\frac{i}{\pi}\right) \cos\left(\frac{|R_z| \omega}{\lambda} \right)\right.\\
& + \left. i\pi \omega\left(\exp\left(i R \omega\left[1+\frac{R_z^2}{2 \lambda^2 R^2}\right]\right)\right.\right.\\
& \left.\left.\times \left[R \omega\left(\frac{R_z^2}{\lambda^2 R^2}-1\right) +i\left( \frac{R_z^4 \omega^2}{8 \lambda^4 R^2}-1\right) \right] \right)\right],
\end{split}
\end{equation}
for $\omega<0$.
After integrating over $\omega$, and retaining the most relevant  asymptotic terms in $R$, we get \cite{suppl}
\begin{align}
	\chi^{S,P}_{x,x} (\bo R, \omega_F) & \simeq -\chi^{S,S}_{x,x}(\bo R,\omega_F) \cos(2\theta_R),\\
	\chi^{S,P}_{z,z} (\bo R, \omega_F) & \simeq \chi^{S,S}_{x,x}(\bo R,\omega_F),\\
	\chi^{S,P}_{x,y} (\bo R, \omega_F) & \simeq -\chi^{S,S}_{x,x}(\bo R, \omega_F) \sin(2\theta_R),
\end{align}
where $\chi^{S,S}_{x,x}$ is given by Eq.\ \eqref{chi_diag}.
Unlike the case of the diagonal orbital components, now there is a strong angular dependence in the in-plane interaction, while the Ising component is angle-independent (see Fig.\ \ref{fig:lattice} for a schematic of high symmetry directions).
We can see that along the $x$ direction, $\theta_R=0$, the DM term vanishes and one ends up with an interaction where the in-plane and Ising terms are out of phase (opposite signs), but with the same magnitude. When $\theta_R = \pi/4$, the in-plane Heisenberg term vanishes so only Ising and DM terms survive, with equal strength and in phase. For separations along the $y$-axis, $\theta_R = \frac{\pi}{2}$, the  DM term vanishes, which produces a completely isotropic Heisenberg interaction, as in the case without orbital mixing. 
There are two other high symmetry directions in the lattice. One corresponds to angles $\theta_R = \pm \frac{\pi}{6}$  (see Fig.\ \ref{fig:lattice}). These angles give prefactors for the different terms: $\frac12$ for XX, and $\pm \frac{\sqrt{3}}{2}$ for DM. This implies that the Ising component dominates over the other two in this direction, its magnitude twice XX, and out of phase with each other. At the same time, the DM term is $\sqrt{3}$ times bigger than the isotropic in-plane (but smaller than Ising), and its sign depends on the specific direction. For $\theta_R = \pm \frac{2 \pi}{3}$, the prefactors are $-\frac12$, $\mp \frac{\sqrt{3}}{2}$ for XX and DM respectively, which makes it similar to the former but with different relative phases.
Other angles (lattice directions) produce interactions that mix all three components, giving a tendency to complex spiral ordering of spins embedded in this lattice.
Exactly at the Dirac point, we find that the decay is even faster than for the diagonal case, $\chi^{S,P} \sim R^{-5}$, and again it does not oscillate.

A likely location for impurities is at the center of the unit cell (Fig.\ \ref{fig:lattice}). There, it preserves inversion symmetry locally. It is probable that each impurity will hybridize to both $S$ and $P$ orbitals, in which case the effective interaction will have contributions from both diagonal and off-diagonal components. In the simple case in which the hybridization to each orbital is of the same magnitude, one can add all the components to obtain the final effective interaction. Given that $\chi^{P,S}_{x,y} = -\chi^{S, P}_{x,y}$, the DM term will vanish, and $\chi_{x,x} = 4\sin^2(\theta_R) \chi^{S,S}_{x,x}$,  $\chi_{z,z} = 4 \chi^{S,S}_{x,x}$. In this case we recover an isotropic interaction for $\theta_R = \frac{\pi}{2}$, and for $\theta_R = 0, \pi$ the interaction is only along the $z$ direction.

\emph{Conclusions.} We have obtained asymptotic expressions for the RKKY interaction in 3D Dirac semimetals. In the limit in which $R\gg R_z$, the indirect coupling decays as $R^{-3}$, where $R$ is the impurity separation in the $xy$ plane. There are three important factors that come into play for the resultant interaction. First, the Fermi velocity anisotropy modifies the period of the oscillation as a function of the impurity separation, and also its magnitude.
Second, the position of the Dirac points in the Brillouin zone, given by $K_z$, results in a second modulation along the $z$ direction, with a period that depends on the $K_z$ value.
Lastly, the orbitals to which the impurities hybridize have impact on the angular dependence of the interaction in the $xy$ plane. When both impurities couple to the same type of orbital ($S$ or $P$), the interaction is angular-independent. When impurities hybridize to a different orbital, there is a strong modulation with the orientation in the lattice. The different components of the interaction survive depending on the directions along the crystal, resulting in complex equilibrium configurations for an impurity ensemble.

\emph{Acknowledgments.} Supported by NSF grant DMR-1508325.

\emph{Note.} During the preparation of this manuscript, we became aware of two related preprints \cite{Chang2015, Hosseini2015} which analyze the RKKY interaction in Weyl semimetals.

\bibliography{refs_3D_Dirac}
\bibliographystyle{apsrev4-1}

\onecolumngrid

\section{Supplemental material for ''Hybridization and anisotropy in the exchange interaction in 3D Dirac semimetals''}

\subsection{Calculation details}

\subsubsection{Diagonal terms}

We start by considering the case in which each impurity is hybridized to the same type of orbital, either $S$ or $P$. In this case we have that $G_0^{\nu=1, S, S}(\bo R) = G_0^{\nu=-1, P, P}(\bo R) = \rho^{-1} (\omega_+ + \lambda q_z)$, and $G_0^{\nu=-1, S, S}(\bo R) = G_0^{\nu=1, P, P}(\bo R) = \rho^{-1} 
(\omega_+ - \lambda q_z)$ [see the main text for notation]. Then, 
\begin{equation} \label{eq:g0ss}
\begin{split}
G_0^{\nu,S,S}(\bs R, \omega) = \frac{1}{(2\pi)^3}\int 
G_0^{\nu,S,S}(\bs\kappa, \omega) e^{i \bo k\cdot\bo R} d\bo k= -\frac{e^{i \nu K_z R_z}}{(2\pi)^2}\!\int_{-\infty}^\infty  e^{i q_z 
	R_z} (\omega_+ +\nu \lambda q_z) K_0\left(\!\!R \sqrt{\lambda^2 
	q_z^2 - \omega_+^2}\right) dq_z,
\end{split}
\end{equation}
with $\bo R$  expressed in cylindrical coordinates, where $R$ is the radial coordinate in the $xy$ plane, and $K_0$ is the modified Bessel function of the second kind. In Eq.\ \eqref{eq:g0ss} we have already integrated over the angle $\theta_k$, and also over $k$ in the $(0,\infty)$ range, which is a valid approximation for large impurity separation. $k_z$ has been replaced by $\pm K_z + q_z$ as well, and we are left with the integration over $q_z$.
Using the fact that 
\begin{equation}
\sqrt{\lambda^2 q_z^2 - \omega_+^2} =
\begin{cases}
\sqrt{\lambda^2 q_z^2 - \omega^2}, & |\omega|\leq \lambda q_z \\
-i\sgn(\omega) \sqrt{\omega^2-\lambda^2 q_z^2}, & |\omega|> 
\lambda 
q_z.
\end{cases}
\end{equation}
we get that 
\begin{equation}
G_0^{\nu,S,S}(\bo R, \omega) = -\frac{e^{i \nu K_z R_z}}{(2\pi)^2}\left(I_{0a}- i\sgn(\omega) I_{0b}\right),
\end{equation}
where
\begin{equation}
\begin{split}
I_{0a} = \left(\int_{-\infty}^{-\frac{|\omega|}{\lambda}}+ 
\int_{\frac{|\omega|}{\lambda}}^{\infty} \right)  e^{i q_z 
	R_z} (\omega +\nu \lambda q_z) K_0\left(\!R \sqrt{\lambda^2 q_z^2 - 
	\omega^2}\right)dq_z,
\end{split}
\end{equation}
and
\begin{equation}
\begin{split}
I_{0b} =  \int_{-\frac{|\omega|}{\lambda}}^{\frac{|\omega|}{\lambda}}
e^{i q_z R_z} (\omega +\nu \lambda q_z)
K_0\left(\!-i\sgn(\omega) R \sqrt{\omega^2 -\lambda^2 q_z^2}\right) dq_z,
\end{split}
\end{equation}
Using the identities
\begin{equation}
K_0(i x) = i \frac{\pi}{2} H_0^{(1)}(x)=i \frac{\pi}{2}[J_0(x) + i Y_0(x)],\qquad (x 
\in \mathbb{R}),
\end{equation}
we get that
\begin{equation}
K_0\left(\!-i\sgn(\omega) R \sqrt{\omega^2 -\lambda^2 q_z^2}\right) = -\frac{\pi}{2} \left[Y_0\left(\!R \sqrt{\omega^2 -\lambda^2 q_z^2}\right)-i \sgn(\omega) J_0\left(\!R \sqrt{\omega^2 -\lambda^2 q_z^2}\right)\right],
\end{equation}
so
\begin{equation}
\begin{split}
I_{0b} =  -\frac{\pi}{2}\int_{-\frac{|\omega|}{\lambda}}^{\frac{|\omega|}{\lambda}}
e^{i q_z R_z} (\omega +\nu \lambda q_z) \left[Y_0\left(\!R \sqrt{\omega^2 -\lambda^2 q_z^2}\right)-i \sgn(\omega) J_0\left(\!R \sqrt{\omega^2 -\lambda^2 q_z^2}\right)\right] dq_z,
\end{split}
\end{equation}
and using the parity properties of the integrand under $q_z \rightarrow 
-q_z$, we can write
\begin{equation}
\begin{split}
I_{0a} =  2\omega \int_{\frac{|\omega|}{\lambda}}^{\infty} \cos\left( 
q_z R_z\right) K_0\left(\!R \sqrt{\lambda^2 q_z^2 - \omega^2}\right) dq_z
+ 2 i \nu \lambda \int_{\frac{|\omega|}{\lambda}}^{\infty} q_z \sin\left( 
q_z R_z\right) K_0\left(\!R \sqrt{\lambda^2 q_z^2 - \omega^2}\right)dq_z,
\end{split}
\end{equation}
and
\begin{equation}
I_{0b} = -\pi \left(I_{0b,1}+I_{0b,2}+I_{0b,3}+I_{0b,4}\right),
\end{equation}
where
\begin{equation}
\begin{split}
I_{0b,1} &= \omega \int_{0}^{\frac{|\omega|}{\lambda}} \cos\left( 
q_z R_z\right) Y_0\left(\!R \sqrt{\omega^2 -\lambda^2 q_z^2}\right)dq_z,\\
I_{0b,2} &= -i|\omega| \int_{0}^{\frac{|\omega|}{\lambda}} \cos\left( 
q_z R_z\right) J_0\left(\!R \sqrt{\omega^2 -\lambda^2 q_z^2}\right)dq_z,\\
I_{0b,3} &= i\nu\lambda \int_{0}^{\frac{|\omega|}{\lambda}}\! q_z 
\sin\left( 
q_z R_z\right) Y_0\left(\!R \sqrt{\omega^2 -\lambda^2 q_z^2}\right)dq_z,\\
I_{0b,4} &= \nu\lambda\sgn(\omega)\! \int_{0}^{\frac{|\omega|}{\lambda}} \!\!
q_z \sin\left(q_z R_z\right) J_0\left(\!R \sqrt{\omega^2 -\lambda^2 
	q_z^2}\right)dq_z.
\end{split}
\end{equation}
These integrals cannot be solved in a closed form. Substituting $u=R\sqrt{\lambda^2 q_z^2- \omega^2}$ in $I_{0a}$,  $v=R\sqrt{\omega^2-\lambda^2 q_z^2}$ in $I_{0b}$, and defining $r=\frac{|R_z|}{\lambda R}$, $\alpha=R|\omega|$, we get
\begin{equation} \label{eq:I0a}
\begin{split}
I_{0a,1} &= \frac{2\omega}{\lambda R} \int_0^\infty 
\frac{u}{\sqrt{\alpha^2+u^2}} \cos\left(\!r\sqrt{\alpha^2+u^2}\right) K_0(u) du,\\
I_{0a,2} &= i\frac{2\nu\sgn(R_z)}{\lambda R^2} \int_0^\infty u\sin\left(\!r\sqrt{\alpha^2+u^2}\right) K_0(u)du,
\end{split}
\end{equation}
and
\begin{equation}
\begin{split}
I_{0b,1} &= \frac{\omega}{\lambda R} \int_0^\alpha 
\frac{v}{\sqrt{\alpha^2-v^2}} \cos\left(\!r\sqrt{\alpha^2-v^2}\right) Y_0(v) dv,\\
I_{0b,2} &= -i\frac{|\omega|}{\lambda R} \int_0^\alpha \frac{v}{\sqrt{\alpha^2-v^2}}\cos\left(
\!r\sqrt{\alpha^2-v^2}\right) J_0(v)dv,\\
I_{0b,3} &= i\frac{\nu\sgn(R_z)}{\lambda R^2} \int_0^\alpha\! v
\sin\left(\!r\sqrt{\alpha^2-v^2}\right) Y_0(v) dv,\\
I_{0b,4} &= \frac{\nu\sgn(\omega R_z)}{\lambda R^2}\! \int_0^\alpha \!\!
v \sin\left(\!r\sqrt{\alpha^2-v^2}\right) J_0(v) dv.
\end{split}
\end{equation}
We  tackle the integrals in the limit $r\ll 1$, expanding the sines and cosines around $r=0$ as
\begin{equation}
\begin{split}
\sin\left(\!r \sqrt{\alpha^2\pm u^2}\right) &= \sum_{n=0}^{\infty}
\frac{(-1)^n}{(2 n+1)!} \left(\!r \sqrt{\alpha^2 \pm u^2}\right)^{2 n+1},\\
\cos\left(\!r \sqrt{\alpha^2 \pm u^2}\right) &=\sum_{n=0}^{\infty}
\frac{(-1)^n}{(2 n)!} \left(\!r \sqrt{\alpha^2 \pm u^2}\right)^{2 n}.
\end{split}
\end{equation}
Then we have that
\begin{equation}
\begin{split}
&\int_0^\infty u \left(\!\sqrt{\alpha^2+u^2}\right)^{2n-1} K_0(u) du = \frac{2^{2n-1}}{\Gamma\left(\frac12 - n\right)} \MeijerG*{3}{1}{1}{3}{n+\frac12}{0,n+\frac12,n+\frac12}{\frac{\alpha^2}{4}},\\
&\int_0^\alpha v \left(\!\sqrt{\alpha^2-v^2}\right)^{2 
	n-1} Y_0(v) dv = 4^n \Gamma\left(\!n+\frac12\right) 
\MeijerG*{2}{1}{2}{4}{n+\frac12,n}{n+\frac12,n+\frac12,0,n}{\frac{\alpha^2}{4}},
\\
&\int_0^\alpha v \left(\!\sqrt{\alpha^2-v^2}\right)^{2 n-1} J_0(v) dv = 2^{n-\frac12} \alpha^{n+\frac12} \Gamma\left(\!n+\frac12\right) J_{n+\frac12}(\alpha),
\end{split}
\end{equation}
where $G^{m,n}_{p,q}$ is the Meijer function.
The summations cannot be done analytically as they are, but we can use asymptotic expansions in $\alpha$
\begin{equation}
\begin{split}
&\MeijerG*{3}{1}{1}{3}{n+\frac12}{0,n+\frac12,n+\frac12}{\frac{\alpha^2}{4}} \simeq 4^{\frac{1}{2}-n} \Gamma\left(\frac{1}{2}-n\right)\alpha^{2n-1},\\
&\MeijerG*{2}{1}{2}{4}{n+\frac12,n}{n+\frac12,n+\frac12,0,n}{\frac{\alpha^2}{4}}
\simeq \frac{4^{-n} \alpha ^n }{\pi }\left(\frac{2 \alpha ^{n-1}}{\Gamma \left(n+\frac{1}{2}\right)}-\sqrt{\pi } 2^n \cos \left(\alpha - \frac{\pi  n}{2} \right)\right),\\
& J_{n+\frac12}(\alpha)\simeq \sqrt{\frac{2}{\pi \alpha}}
\sin  \left(\alpha -\frac{\pi  n}{2}\right).
\end{split}
\end{equation}
Now the summation can be performed, and we get
\begin{equation}
\begin{split}
\Sigma_{0a,1}(\alpha,r) = &\sum_{n=0}^\infty \frac{(-1)^n \alpha^{2n-1} r^{2 n} }{(2n)!} = \frac{\cos(r\alpha)}{\alpha},\\
\Sigma_{0a,2}(\alpha,r) = &\sum_{n=0}^\infty \frac{(-1)^n \alpha^{2n+1} r^{2n+1} }{(2n+1)!} = \sin(r\alpha),\\
\Sigma_{0b,1}(\alpha,r) = &\sum_{n=0}^\infty \frac{(-1)^n \ r^{2 n}}{(2 n)!} \left[\frac{2\alpha^{2n-1}}{\pi}-\frac{2^n \alpha^n}{\sqrt{\pi}}\Gamma\left(\!n+\frac12\right)\cos\left(\alpha-\frac{n\pi}{2}\right)\right]= \frac{2}{\pi\alpha}\cos(r\alpha) - \cos\left(\alpha\left[1+\frac{r^2}{2}\right]\right), \\
\Sigma_{0b,2}(\alpha,r) = &-\sqrt{\frac{2}{\pi \alpha}}\sum_{n=0}^\infty \frac{(-1)^n  r^{2 n} }{(2 	n)!} 2^{n-\frac{1}{2}} \alpha ^{n+\frac{1}{2}} \Gamma \left(\!n+\frac{1}{2}\right)\sin  \left(\!\alpha -\frac{\pi  n}{2}\right)=\sin\left(\alpha\left[1+\frac{r^2}{2}\right]\right),\\
\Sigma_{0b,3}(\alpha,r) = &\sum_{n=0}^\infty \frac{(-1)^n  r^{2n+1}}{(2 n+1)!} \frac{\alpha^{n+1}}{\pi 2^{2n+1} }\left(\frac{\alpha^{n}}{\Gamma \left(n+\frac{3}{2}\right)}-\sqrt{\pi } 2^n \sin \left(\alpha - \frac{\pi  n}{2} \right)\right)=\frac{2}{\pi}\sin(r\alpha)-r\alpha\sin\left(\alpha\left[1+\frac{r^2}{2}\right]\right),\\
\Sigma_{0b,4}(\alpha,r) = &-\sqrt{\frac{2}{\pi \alpha}}\sum_{n=0}^\infty \frac{(-1)^n  r^{2 n+1}}{(2n+1)!} 2^{n+\frac{1}{2}} \alpha^{n+\frac{3}{2}} \Gamma \left(\!n+\frac{3}{2}\right)\cos  \left(\!\alpha -\frac{\pi  n}{2}\right)=-r\alpha\cos\left(\alpha\left[1+\frac{r^2}{2}\right]\right).
\end{split}
\end{equation}
Then
\begin{equation}
\begin{split}
I_{0a,1} &\simeq \frac{2\omega}{\lambda r}\Sigma_1(\alpha, r) =\frac{2\omega \cos(r\alpha)}{\lambda R \alpha} = \frac{2\sgn(\omega)}{\lambda R^2} \cos\left(\frac{|R_z \omega|}{\lambda}\right),\\
I_{0a,2} &\simeq i\frac{2\nu\sgn(R_z)}{\lambda R^2}\Sigma_2(\alpha, r) =i\frac{2\nu\sgn(R_z)}{\lambda R^2}\sin(r\alpha) = i\frac{2\nu\sgn(R_z)}{\lambda R^2}\sin\left(\frac{|R_z \omega|}{\lambda}\right),
\end{split}
\end{equation}
which results in
\begin{equation}
I_{0a}(R,R_z,\omega,\nu) \simeq \frac{2}{\lambda R^2}\sgn(\omega) \exp\left(i\nu\frac{\omega R_z}{\lambda}\right),\quad|Rz|\ll \lambda R.
\end{equation}
On the other hand
\begin{equation}
\begin{split}
I_{0b,1} &\simeq \frac{\omega}{\lambda R} \Sigma_{0b,1}(\alpha,r)=\frac{\omega}{\lambda R}\left[\frac{2}{\pi R |\omega|}\cos\left(\frac{|\omega R_z|}{\lambda}\right)-\cos\left(\!R|\omega|\left[1+\frac{R_z^2}{2 \lambda^2 R^2}\right]\right)\right],\\
I_{0b,2} &\simeq -i\frac{|\omega|}{\lambda R} \Sigma_{0b,2}(\alpha,r)=-i\frac{|\omega|}{\lambda R} \sin\left(\!R|\omega|\left[1+\frac{R_z^2}{2 \lambda^2 R^2}\right]\right) ,\\
I_{0b,3} &\simeq i\frac{\nu\sgn(R_z)}{\lambda R^2} \Sigma_{0b,3}(\alpha,r) = i\frac{\nu\sgn(R_z)}{\lambda R^2} \left[\frac{2}{\pi}\sin\left(\frac{|\omega R_z|}{\lambda}\right)-\frac{|\omega R_z|}{\lambda}\sin\left(\!R|\omega|\left[1+\frac{R_z^2}{2 \lambda^2 R^2}\right]\right)\right] ,\\
I_{0b,4} &\simeq \frac{\nu\sgn(\omega R_z)}{\lambda R^2} \Sigma_{0b,4}(\alpha,r)=
-\frac{\nu  R_z \omega}{\lambda^2 R^2} \cos\left(\!R|\omega|\left[1+\frac{R_z^2}{2 \lambda^2 R^2}\right]\right).
\end{split}
\end{equation}
After some algebra, one finds that
\begin{equation}
I_{0b}(R,R_z,\omega,\nu) = -\frac{\pi}{\lambda R}\sgn(\omega) \left[\frac{2}{\pi R} \exp\left(\! i\nu\frac{\omega R_z} {\lambda}\right)-|\omega|\left(1+\nu\frac{R_z}{\lambda R}\right) \exp\left(\! i R \omega\left[1+\frac{R_z^2}{2\lambda^2 R^2}\right]\right)\right],\quad|Rz|\ll \lambda R.
\end{equation}
For $\omega<0$, this gives rise to the Green's function appearing in Eq.\ (15) in the main text.

\subsubsection{Cross terms}

When one of the impurities hybridizes with an $S$ orbital and the other one with a $P$ orbital, we have that $G_0^{\nu,S,P}(\kappa,\omega)= -G_0^{\nu,P,S}(\kappa,\omega) = i\rho^{-1} k \sin\theta_k$. 

The Fourier transform gives
\begin{equation}
\begin{split}
G^{\nu, S, P}_0(\bo R, \omega) = \frac{i}{(2\pi)^3} \int_{-\infty}^\infty 
dk_z\: e^{i k_z R_z} \int_0^\infty dk\: 
\frac{k^2}{\rho}
\int_0^{2\pi} d\theta_k\: \sin\theta_k\: e^{i k R \cos (\theta_k - 
	\theta_R)}.
\end{split}
\end{equation}
Integrating over $\theta_k$, we get
\begin{equation}
\begin{split}
G_0^{\nu,S,P}(\bo R, \omega) = -\frac{\sin(\theta_R)}{4\pi^2}
\int_{-\infty}^\infty dk_z e^{i k_z R_z}\!
\int_0^\infty\! dk \frac{k^2}{\rho} J_1(k R).
\end{split}
\end{equation}
Integrating over $k$ and replacing $k_z$ by $K^\tau_z + q_z$, we get
\begin{equation}
\begin{split}
G_0^{\nu,S,P}(\bo R,\omega) = \frac{\sin(\theta_R)}{4\pi^2}\:
e^{i K^\tau_z R_z}\int_{-\infty}^\infty dq_z e^{i q_z R_z}
\sqrt{\lambda^2 q_z^2 - \omega_+^2}\: K_1\left(R \sqrt{\lambda^2 
	q_z^2 	- \omega_+^2}\right),
\end{split}
\end{equation}
which can be written as
\begin{equation}
G_0^{\nu,S,P}(\bo R,\omega) =  \frac{\sin(\theta_R)}{4\pi^2}  e^{i 
	K^\nu_z R_z} \left(I_2-i\sgn(\omega) I_3\right),
\end{equation}
where
\begin{equation}
\begin{split}
I_2 =  \left(\int_{-\infty}^{-\frac{|\omega|}{\lambda}} + 
\int_{\frac{|\omega|}{\lambda}}^\infty\right) e^{i q_z R_z}
\sqrt{\lambda^2 q_z^2 - \omega^2} K_1\left(R \sqrt{\lambda^2 q_z^2 - \omega^2} \right) dq_z,
\end{split}
\end{equation}
and
\begin{equation}
\begin{split}
I_3 =  \int_{-\frac{|\omega|}{\lambda}}^{\frac{|\omega|}{\lambda}}
e^{i q_z R_z}  \sqrt{\omega^2-\lambda^2 q_z^2} K_1\left(-iR\sgn(\omega) 
\sqrt{\omega^2 - \lambda^2 q_z^2} \right) dq_z.
\end{split}
\end{equation}
Similarly,
\begin{equation}
G_z^{\nu,S,P}(\bo R,\omega) =  \frac{i \cos(\theta_R)}{4\pi^2}\:  e^{i 
	K^\nu_z R_z} \left(I_2 - i\sgn(\omega) I_3\right)
\end{equation}
with $G_z^{\nu,P,S}(\bo R,\omega) = G_z^{\nu,S,P}(\bo R,\omega)$.
Changing variables as in the previous case, we get
\begin{equation}
I_2 =  \frac{2}{\lambda R^2} \int_0^\infty \frac{u^2 K_1(u)}{\sqrt{u^2 + R^2 
		\omega^2}} \cos\left(\frac{R_z}{\lambda R}\sqrt{u^2 + R^2 
	\omega^2}\right) du.
\end{equation}
This expression is even under $\omega\ra -\omega$ and $R_z \ra -R_z$, then
\begin{equation}
I_2 =  \frac{2}{\lambda R^2} \int_0^\infty \frac{u^2 
	K_1(u)}{\sqrt{u^2 + \alpha^2}} \cos\left(r\sqrt{u^2 + \alpha^2}\right) du.
\end{equation}
Similarly, for $I_3$ we have
\begin{equation}
I_3 =  -\frac{\pi}{\lambda R^2} \int_0^\alpha \frac{v^2 
	H^{(1)}_1\left(\sgn(\omega) v\right)}{\sqrt{\alpha^2 - v^2}} \cos\left(r\sqrt{\alpha^2 - v^2}\right) dv = 
-\frac{\pi}{\lambda R^2} \int_0^\alpha \frac{v^2 
	\left[i Y_1(v) + \sgn(\omega) J_1(v)\right]}{\sqrt{\alpha^2 - v^2}} \cos\left(r\sqrt{\alpha^2 - v^2}\right) dv,
\end{equation}
or 
\begin{equation}
I_3 =- \frac{\pi}{\lambda R^2}\left(\sgn(\omega) I_{3a} + i I_{3b}\right),
\end{equation}
where
\begin{align}
I_{3a} &= \int_0^\alpha \frac{v^2 J_1(v)}{\sqrt{\alpha^2 - v^2}} \cos\left(r\sqrt{\alpha^2 - v^2}\right) dv, \\
I_{3b} &= \int_0^\alpha \frac{v^2 Y_1(v)}{\sqrt{\alpha^2 - v^2}} \cos\left(r\sqrt{\alpha^2 - v^2}\right) dv.
\end{align}
Using the cosine series expansions, we get
\begin{align}
& \int_0^\infty u^2 K_1(u) \left(u^2 + \alpha^2\right)^{n-\frac12} du = \frac{2^{2n-3} \alpha^3}{\Gamma\left(\frac12 - n\right)} \MeijerG*{3}{1}{1}{3}{n - 1}{-\frac32, n-1, n}{\frac{\alpha^2}{4}},\\
& \int_0^\alpha v^2 J_1(v) \left(\alpha^2 - v^2\right)^{n-\frac12} dv = 2^{n-\frac12}\Gamma\left(n+\frac12\right)\alpha^{n+\frac32} 
J_{n+\frac32}(\alpha),\\
& \int_0^\alpha v^2 Y_1(v) \left(\alpha^2 - v^2\right)^{n-\frac12} dv = 2^{n-\frac12}\Gamma\left(n+\frac12\right)\alpha^{n+\frac32} 
Y_{n+\frac32}(\alpha) + \frac{4}{\pi}\alpha^{2n-1}.
\end{align}
The summations cannot be performed analytically, so we expand the special functions for $\alpha\gg 1$
\begin{align}
& \MeijerG*{3}{1}{1}{3}{n - 1}{-\frac32, n-1, n}{\frac{\alpha^2}{4}} \simeq 4^{2-n} \alpha ^{2 (n-2)} \Gamma\left(\frac{1}{2}-n\right),\\
& J_{n+\frac32}(\alpha) \simeq -\sqrt{\frac{2}{\pi \alpha}} \cos 
\left(\alpha-\frac{\pi n}{2}\right) +\frac{(n+1)(n+2) }{\sqrt{2 \pi } \:
	\alpha^{3/2}}\sin \left(\alpha-\frac{\pi  n}{2} \right),\\
&Y_{n+\frac32}(\alpha) \simeq -\sqrt{\frac{2}{\pi \alpha}} \sin 
\left(\alpha-\frac{\pi n}{2}\right)-\frac{(n+1)(n+2) }{\sqrt{2 \pi } \:
	\alpha^{3/2}}\cos \left(\alpha-\frac{\pi  n}{2} \right).
\end{align}
Proceeding with the summations, we get
\begin{align}
\Sigma_2(\alpha, r) =& 2\sum_{n=0}^\infty \frac{(-1)^n}{(2n)!} r^{2n}\alpha^{3+2(n-2)} = \frac{2\cos(r\alpha)}{\alpha},\\
\Sigma_{3a}(\alpha, r) = &-\frac{1}{\sqrt{\pi}}\sum_{n=0}^\infty \frac{(-1)^n}{(2n)!} r^{2n} 2^{n-1} \alpha^n \Gamma \left(n+\frac{1}{2}\right) \left[(n+1) (n+2) \sin \left(\frac{\pi  n}{2}-\alpha \right)+2 \alpha  \cos \left(\frac{\pi  n}{2}-\alpha \right)\right],\\
= & \alpha (r^2-1) \cos\left(\alpha+\frac{\alpha r^2}{2}\right)
-\left(\frac{\alpha^2 r^4}{8}-1\right)\sin\left(\alpha+\frac{\alpha r^2}{2}\right)\\
\Sigma_{3b}(\alpha, r) = &-\sum_{n=0}^\infty \frac{(-1)^n}{(2n)!} r^{2n} \left(2^{n-1} \alpha^{n+\frac32} \Gamma \left(n+\frac{1}{2}\right) \left[\sqrt{\frac{2}{\pi \alpha}} \sin 
\left(\alpha-\frac{\pi n}{2}\right)+\frac{(n+1)(n+2) }{\sqrt{2 \pi } \:
	\alpha^{3/2}}\cos \left(\alpha-\frac{\pi  n}{2} \right)\right] - \frac{4}{\pi}\alpha^{2n-1}\right),\\
= & \alpha (r^2-1)\sin\left(\alpha+\frac{\alpha r^2}{2}\right) +\left(\frac{\alpha^2 r^4}{8}-1\right)\cos\left(\alpha+\frac{\alpha r^2}{2}\right) + \frac{4}{\pi \alpha}\cos(r \alpha).
\end{align}
Then
\begin{align}
I_2 &\simeq \frac{4}{\lambda R^3 |\omega|} \cos\left(\frac{|R_z \omega|}{\lambda} \right),
\end{align}
and
\begin{align}
I_{3a} & \simeq  R|\omega| \left(\frac{R_z^2}{\lambda^2 R^2}-1\right) \cos\left(R \omega\left[1+\frac{R_z^2}{2 \lambda^2 R^2}\right]\right)
-\left(\frac{\omega^2 R_z^4}{8 \lambda^4 R^2}-1\right)\sin\left(R |\omega|\left[1+\frac{R_z^2}{2 \lambda^2 R^2}\right]\right),\\
I_{3b} & \simeq R|\omega| \left(\frac{R_z^2}{\lambda^2 R^2}-1\right) \sin\left(R |\omega|\left[1+\frac{R_z^2}{2 \lambda^2 R^2}\right]\right)
+\left(\frac{\omega^2 R_z^4}{8 \lambda^4 R^2}-1\right)\cos\left(R \omega\left[1+\frac{R_z^2}{2 \lambda^2 R^2}\right]\right) + \frac{4}{\pi R |\omega|} \cos\left(\frac{|R_z \omega|}{\lambda} \right),
\end{align}
which gives
\begin{equation}
I_3 \simeq -\frac{\pi}{\lambda R^2} \left[\exp\left(i R \omega\left[1+\frac{R_z^2}{2 \lambda^2 R^2}\right]\right)
\left[R \omega\left(\frac{R_z^2}{\lambda^2 R^2}-1\right) +i\left( \frac{R_z^4 \omega^2}{8 \lambda^4 R^2}-1\right) \right] + \frac{4i}{\pi R |\omega|}\cos\left(\frac{|R_z \omega|}{\lambda}\right)\right].
\end{equation}
Defining $f(R, R_z, \omega) = I_2(R, R_z, \omega) - i \sgn(\omega) I_3(R, R_z, \omega)$, we have (summing over valleys) that
\begin{align}
G_0^{S,P}(\bo R,\omega) &\simeq  \frac{\sin(\theta_R)}{2\pi^2} \cos\left( 
K_z R_z\right)\: f(R, R_z, \omega),\\
G_z^{S,P}(\bo R,\omega) &\simeq  \frac{i \cos(\theta_R)}{2\pi^2}\cos\left( 
K_z R_z\right)\: f(R, R_z, \omega),
\end{align}
with
\begin{equation}
\begin{split}
f(R, R_z, \omega) =& \frac{1}{\lambda R^2} \left(\frac{4}{R |\omega|} \cos\left(\frac{|R_z \omega|}{\lambda} \right) + i\pi\sgn(\omega)\left[\exp\left(i R \omega\left[1+\frac{R_z^2}{2 \lambda^2 R^2}\right]\right)
\left[R \omega\left(\frac{R_z^2}{\lambda^2 R^2}-1\right) +i\left( \frac{R_z^4 \omega^2}{8 \lambda^4 R^2}-1\right) \right]\right.\right.\\
& +\left.\left. \frac{4i}{\pi R |\omega|}\cos\left(\frac{|R_z \omega|}{\lambda}\right) \right]\right),
\end{split}
\end{equation}
and, for $\omega<0$,
\begin{equation}
\begin{split}
f(R, R_z, \omega<0) =& -\frac{1}{\lambda R^2} \left(\frac{4}{R \omega} \cos\left(\frac{|R_z| \omega}{\lambda} \right) + i\pi \omega\left[\exp\left(i R \omega\left[1+\frac{R_z^2}{2 \lambda^2 R^2}\right]\right)
\left[R \omega\left(\frac{R_z^2}{\lambda^2 R^2}-1\right) +i\left( \frac{R_z^4 \omega^2}{8 \lambda^4 R^2}-1\right) \right]\right.\right.\\
& + \left.\left. \frac{4i}{\pi R \omega}\cos\left(\frac{|R_z| \omega}{\lambda}\right) \right]\right).
\end{split}
\end{equation}
Integrating in $\omega$, and to lowest order in $1/R$, one gets
\begin{align}
\chi^{S,P}_{x,x} &\simeq \frac{\omega_F^2}{4 \pi^3 \lambda^2 R^3} \cos^2\left(K_z R_z\right) \cos(2\theta_R) \cos\left(2 R \omega_F \left[1+\frac{R_z^2}{2 \lambda^2 R^2}\right]\right),\\
\chi^{S,P}_{z,z} &\simeq -\frac{\omega_F^2}{4 \pi^3 \lambda^2 R^3} \cos^2\left(K_z R_z\right)  \cos\left(2 R \omega_F \left[1+\frac{R_z^2}{2 \lambda^2 R^2}\right]\right),\\
\chi^{S,P}_{x,y} &\simeq -\frac{\omega_F^2}{4 \pi^3 \lambda^2 R^3} \cos^2\left(K_z R_z\right) \cos(2\theta_R) \sin\left(2 R \omega_F \left[1+\frac{R_z^2}{2 \lambda^2 R^2}\right]\right).
\end{align}
Then
\begin{equation}
H^{S,P}_{\text{RKKY}} = \frac{J^2 \omega_F^2}{4 \pi^3 \lambda^2 R^3} \cos^2\left(K_z R_z\right) \cos\left(2 R \omega_F \left[1+\frac{R_z^2}{2 \lambda^2 R^2}\right]\right) \left[\cos(2\theta_R) \bs S_1\cdot\bs S_2 - 2\cos^2(\theta_R) S_1^z S_2^z - \sin(2\theta_R) (\bs S_1\times\bs S_2)_z\right].
\end{equation}

\end{document}